\documentclass[%
reprint,twocolumn,
preprintnumbers,
nofootinbib,
amsmath,amssymb,
aps,
prl,
]{revtex4-2}

\usepackage{preamble}
\begin{document}
\preprint{FERMILAB-PUB-25-0407-PPD-T}

\title{The isotropy of Hubble expansion in the early and late Universe}

\author{Alan Junzhe Zhou}
\affiliation{
    Department of Physics, University of Chicago, Chicago, IL 60637, USA
}
\affiliation{
    Kavli Institute for Cosmological Physics, University of Chicago, Chicago, IL 60637, USA}
\author{Scott Dodelson}
\affiliation{
    Fermi National Accelerator Laboratory, P.O. Box 500, Batavia, IL 60510, USA}
\affiliation{
    Department of Astronomy and Astrophysics, University of Chicago, Chicago, IL 60637, USA
}
\affiliation{
    Kavli Institute for Cosmological Physics, University of Chicago, Chicago, IL 60637, USA
}
\author{Daniel Scolnic}
\affiliation{
    Department of Physics, Duke University, Durham, NC 27708, USA
}

\begin{abstract}
We test the isotropy of Hubble expansion by combining several probes for the first time, constructing full-sky maps of expansion rate variation using Type Ia supernovae, fundamental plane galaxies, and CMB temperature fluctuations. We find no hint of anisotropy or correlation between early- and late-Universe expansion across all systematic models. The 99\% confidence upper limits on expansion rate anisotropy are 0.39\% for low-redshift supernovae, 0.95\% for high-redshift CMB, and 0.37\% when combined at a 60-degree smoothing scale. A significant anomaly in the fundamental plane residual map may reflect systematics in the current DESI dataset, as evidenced by the absence of cross-correlation with other tracers and its correlation with spatial density variations.
\end{abstract}

\maketitle
\section{Introduction}
The standard $\Lambda$CDM cosmological model assumes homogeneity and isotropy: that the laws of physics are invariant throughout the Universe and that the distribution of matter and energy is statistically uniform on large scales. A direct consequence is an isotropic cosmic expansion. The expansion history, $H(z)$, and its present-day value, the Hubble constant $H_0 = 100h$~km\,s$^{-1}$\,Mpc$^{-1}$, are constrained by early-Universe temperature anisotropies of the Cosmic Microwave Background (CMB) and by late-Universe distance indicators such as Type Ia supernovae (SNe) and fundamental plane (FP) galaxies.

Yet two decades of CMB observations~\cite{
hinshawFirstYearWilkinson2003, 
spergelFirstYearWilkinsonMicrowave2003, 
deoliveira-costaSignificanceLargestScale2004, 
eriksenPointCorrelationFunctions2005, 
landAxisEvil2005, 
copiLargeangleAnomaliesCMB2010, 
bennettSevenYearWilkinsonMicrowave2011, 
planckcollaborationPlanck2018Results2020e} have revealed hints of statistical anomalies in the low-$\ell$ temperature multipoles. Low-redshift surveys have also raised questions about the nature of the recent cosmic acceleration \cite{collaborationDESI2024VI2025, collaborationDESIDR2Results2025}. Furthermore, $h$ inferred from early-Universe CMB data remains in tension with that from local supernovae measurements \cite{valentinoRealmHubbleTension2021, kamionkowskiHubbleTensionEarly2022}. A non-uniform Hubble flow could potentially alleviate these tensions through scenarios such as vector field inflation, non-trivial topologies, or inhomogeneous dark energy \cite{
deoliveira-costaSignificanceLargestScale2004,
copelandDynamicsDarkEnergy2006,
coorayMeasuringDarkEnergy2010,
maleknejadGaugeFieldsInflation2013
}. However, the scarcity of independent large-scale modes makes it difficult to quantify such anomalies precisely. 

We test for anisotropy in the Hubble flow by combining information from both early- and late-Universe probes for the first time. Under $\Lambda$CDM, any directional variation in $H_0$ measurements should be \emph{consistent with statistical noise and sample variance} and also \emph{uncorrelated between epochs}. A significant violation of this null hypothesis would suggest new physics or previously unaccounted-for systematics. While Hubble flow isotropy has been examined in individual probes~\citep{
colinProbingAnisotropicLocal2011, 
kimLimitsAnisotropicInflation2013,
saadehHowIsotropicUniverse2016,  
soltisPercentLevelTestIsotropic2019,  
fosalbaExplainingCosmologicalAnisotropy2021,
Cowell_2022, 
yeungDirectionalVariationsCosmological2022,
haridasuRadialTullyFisherRelation2024, 
gimeno-amoExploringStatisticalIsotropy2025}, we systematically construct full-sky $h$ maps using SNe, FP, and CMB data and compare them with isotropic predictions. The SN and FP maps trace the local expansion ($z < 1$), whereas the CMB map extrapolates the expected $h$ from the $z\approx1100$ temperature fluctuations. Not only do they independently probe the isotropy of the low- and high-$z$ Hubble flow, but their cross-correlation also provides a robust test for any correlated signal between the early and late Universe.

\section{Hubble residual maps}
\label{sec:map}
\begin{figure*}[htbp]
    \centering
    \begin{minipage}[b]{0.32\linewidth}
        \includegraphics[width=\linewidth]{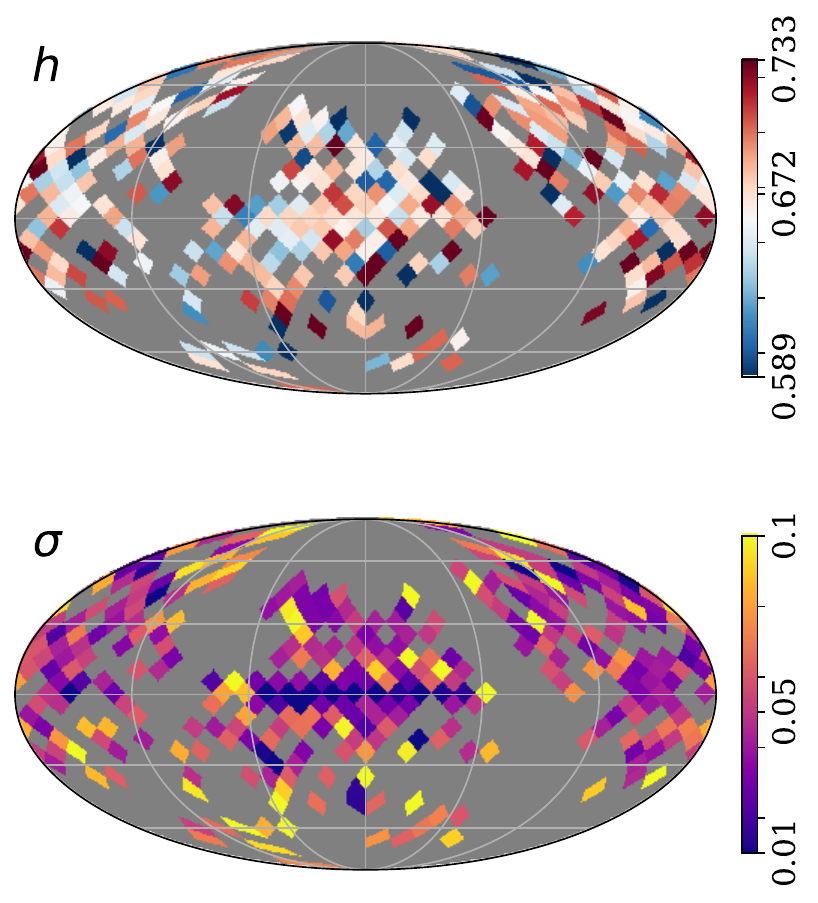}
    \end{minipage}
    \hfill
    \begin{minipage}[b]{0.32\linewidth}
        \includegraphics[width=\linewidth]{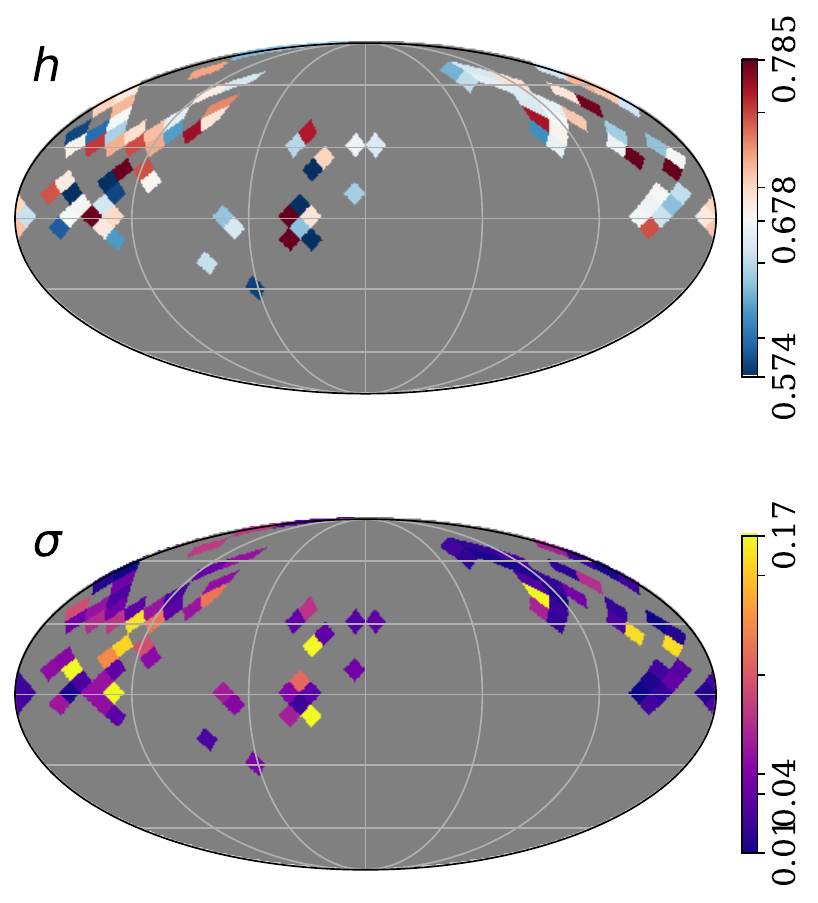}
    \end{minipage}
    \hfill
    \begin{minipage}[b]{0.32\linewidth}
        \includegraphics[width=\linewidth]{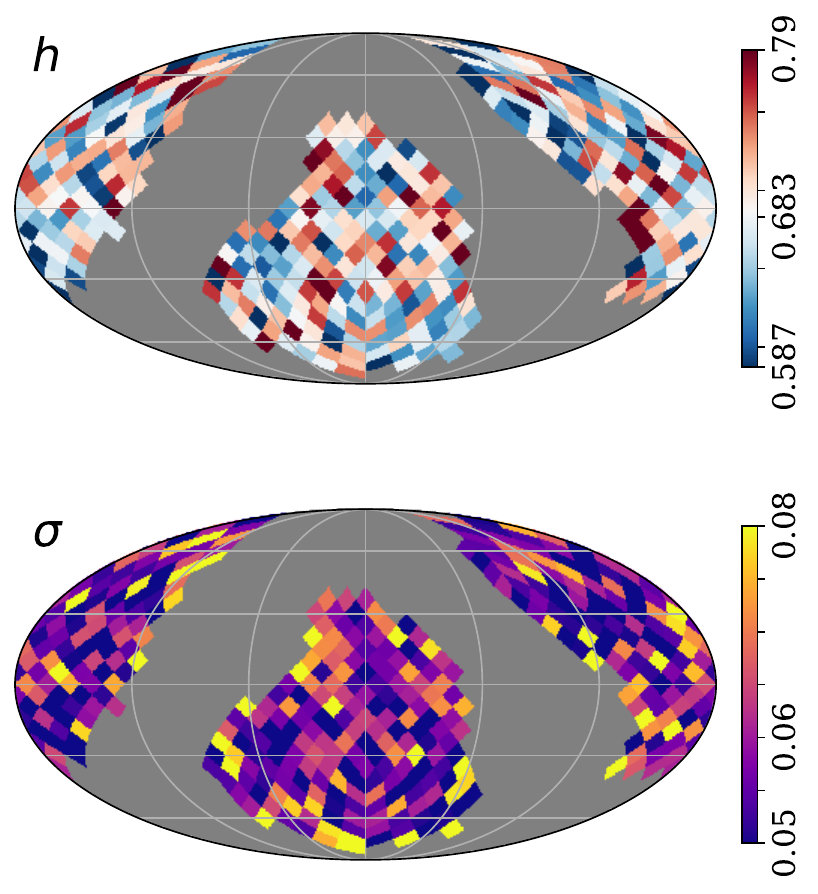}
    \end{minipage}
    \caption{The $h$ and uncertainty $\sigma$ maps for \textbf{supernovae} (left), \textbf{fundamental plane galaxies} (middle) and \textbf{CMB} (right). All maps are calibrated to the Planck cosmology using fiducial analysis settings. 
    Color bars mark the 5th, 50th and 95th percentiles. $\fsky = 0.41, 0.12, 0.52$.
    }
    \label{fig:maps}
\end{figure*}

\begin{figure}
    \centering
    \includegraphics[width=1.0\linewidth]{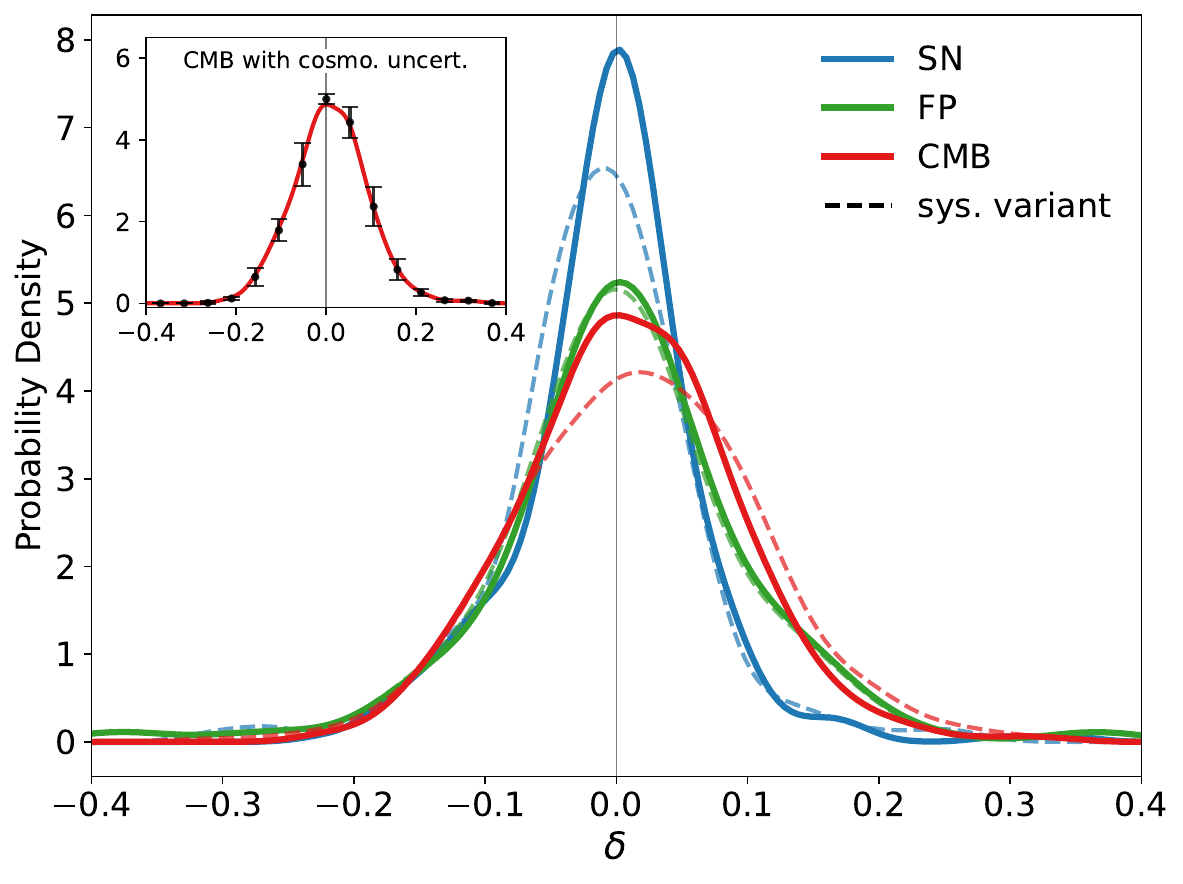}
    \caption{Pixel value histograms of $\dsn$, $\dfp$, and $\dcmb$. Solid lines show fiducial configurations; dashed lines show systematic variants. The inset shows $\dcmb$ histograms with error bars indicating 1$\sigma$ cosmological uncertainty. The distributions are zero-centered and approximately Gaussian. $\dsn$ has the smallest scatter and $\dcmb$ the largest.}
    \label{fig:hists}
\end{figure}

\begin{figure*}
    \centering
    \includegraphics[width=1.0\linewidth]{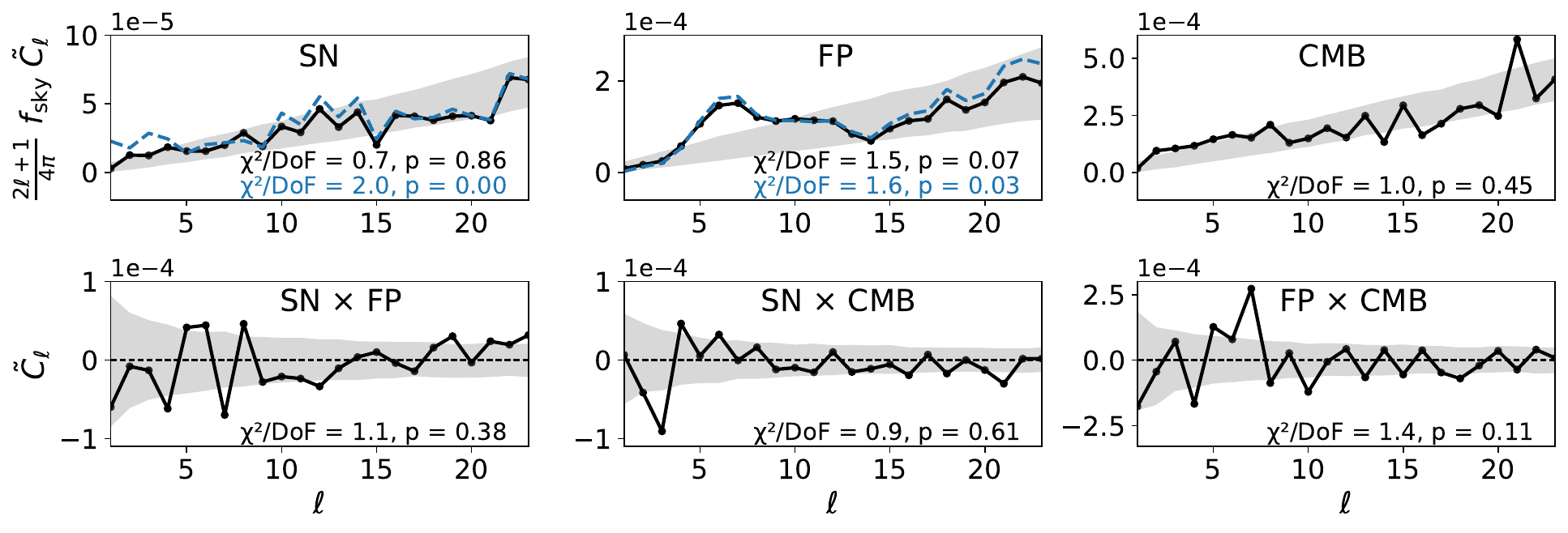}
    \caption{Weighted $C_\ell$ (\Eq\ref{eqn:cl}) of the Hubble residual maps $\delta_{\rm SN}$, $\delta_{\rm FP}$, and $\delta_{\rm CMB}$. The weighted auto-$C_\ell$ is scaled by $\fsky$ so that the area under the curve approximates the variance of the full-sky $\delta$ field. Solid lines represent the data, while shaded regions represent the 1$\sigma$ intervals of isotropic noise realizations. For the fiducial model (black), $\dsn$ and $\dcmb$ are consistent with isotropy, whereas the FP auto-$C_\ell$ is not, with a significant peak around $\ell = 6$. The SN and FP systematic variants (blue) do not remove peculiar velocities and hence exhibit anisotropic contamination from the large-scale structures.}
    \label{fig:cl_cross}
\end{figure*}

{\noindent\it Supernovae.} Type Ia supernovae (SNe) are precise standardizable candles for cosmological distance measurements. The primary observable is the distance modulus $\mu = m - M$, where $m$ and $M$ are the apparent and standardized absolute magnitudes, respectively. For a supernova $i$ at scale factor $a_i$ (redshift $z_i$), the theoretical distance modulus $\mu_{\text{th},i}$ for a flat universe is
\begin{align}
    \mu_{\text{th},i} 
    \label{eqn:mu_th}
    &= 5 \log_{10}\left[\frac{1}{\mathrm{Mpc}} \left(\frac{c}{a_i}  \int_{a_i}^1  \frac{d a'}{a'^2 H(a')}\right)\right] + 25 \,,
\end{align}
where the argument of the logarithm is the luminosity distance $d_L$ in Mpc. 
Equation~\ref{eqn:mu_th} assumes an isotropic Hubble flow. We hypothesize that the difference between the observed $\mu_i$ and $\mu_{\text{th},i}$ arises from noise and potential anisotropy in the cosmic expansion. We model the latter as a perturbation to the Hubble parameter along the line of sight to each supernova
\begin{equation}
    \label{eqn:hubble}
    H_{0,i} = H_0 (1 + \delta_i) \,.
\end{equation}
In the limit of noiseless measurement, the distance modulus residual is directly proportional to the Hubble flow perturbation \cite{soltisPercentLevelTestIsotropic2019}, since 
\begin{align}
    \mu_i - \mu_{\text{th},i}
    &= 5 \log_{10} \left( \int_{a_i}^1  \frac{d a'}{a'^2 H_i(a')} \Big/ \int_{a_i}^1  \frac{d a'}{a'^2 H(a')} \right) \\
    \label{eqn:delta_sn}
    &= - 5 \log_{10} \left( 1 + \delta_i \right) \approx - \frac{5 \, \delta_i}{\ln 10} \,.
\end{align}
Equation~\ref{eqn:delta_sn} allows us to estimate $\delta_i$ and its uncertainty $\sigma_i$ from each supernova. Next, we directly construct a full sky $\delta$ map using the \texttt{nside} = 8 HEALPix pixelization scheme \cite{gorskiHEALPixFrameworkHigh2005,zoncaHealpyEqualArea2019} in contrast to the more indirect normalized residual approach of \citet{soltisPercentLevelTestIsotropic2019}. The value and uncertainty of each pixel $p$ is given by the unbiased inverse variance estimator 
\begin{equation}
    \label{eqn:inv_var}
    \delta_\text{SN}(p) = \frac{\sum_{i\in p} \delta_i / \sigma_i^2}{\sum_{i\in p} 1/ \sigma_i^2} \,, \quad \sigma_\text{SN}(p) = \frac{1}{\sqrt{\sum_{i \in p} 1/\sigma_i^2}} \,.
\end{equation}

We analyze the Pantheon+SH0ES SNe data from \citet{riessComprehensiveMeasurementLocal2022},
originally calibrated via Cepheids to the absolute magnitude $M_\text{fid} = -19.24$ and yielding a global measurement of $h = 0.7330$. Since we focus on the anisotropy rather than the absolute value of $h$ and wish to compare with other distance tracers, we recalibrate the SNe to the Planck 2018 CMB cosmology ($h = 0.6736$ \cite{planckcollaborationPlanck2018Results2020a}), thereby changing the absolute magnitude 
to $M = -19.42$. The recalibration only centers the distribution of $h$ around $h_\text{Planck}$ but has no impact on our anisotropy results. We include 1701 SN with $z<2.3$ and median $z=0.164$. The $h$ and uncertainty maps are shown in \Fig\ref{fig:maps}, and the pixel histogram in \Fig\ref{fig:hists}. 
The relative motion between the SNe and the observer modifies the SNe redshift and coherently distorts of $\delta_\text{SN}$ \cite{Hui_2006, Huterer_2017}.
In the fiducial analysis, we consider supernova redshifts corrected for both the CMB dipole and peculiar velocities due to local group and host motions \cite{riessComprehensiveMeasurementLocal2022}. The systematic variant where the redshift ($z_\text{CMB}$) is corrected only for the former is shown in Fig.~2 as the blue dashed curve. 
\\\\
{\noindent\it Fundamental Plane Galaxies.}
Empirically, the physical radius of elliptical galaxies is related to their dynamical properties, and can thus be calibrated as a standardized ruler for distance and expansion rate measurements. 
This relationship, called the fundamental plane (FP), relates a galaxy's effective radius $R_e$, central velocity dispersion $\sigma$, and mean surface brightness within $R_e$, $I_e$, 
\begin{equation}
    \log R_e = a \log \sigma + b I_e + c \,.
\end{equation}
We analyze the DESI FP galaxy data~\cite{saidDESIPeculiarVelocity2024},
where $a$ and $b$ are calibrated to the observed galaxy population, and the zero-point $c$ is calibrated to the surface brightness fluctuation distance of 
the Coma cluster \cite{saidDESIPeculiarVelocity2024, jensenInfraredSurfaceBrightness2021}. 
We estimate each galaxy's angular diameter distance $d_A$ (Mpc) by combining the FP-inferred physical radius $R_e$ (kpc) with the observed angular size $\theta_e$ (arcseconds),
\begin{equation}
    k_i \equiv \log_{10} d_{A,i} = \log_{10}\left( \frac{R_{e,i}}{\theta_{e,i}} \cdot \frac{648}{\pi} \right)\,,
\end{equation}
where $d_{A,i}$ is the angular diameter distance
.

Analogous to the SNe analysis, if the expansion rate in direction $i$ deviates from the homogeneous background by a perturbation $\delta_i$ (\Eq\ref{eqn:hubble}), the observed angular diameter distance differs from the isotropic theory by
\begin{align}
    k_i - k_{\rm th,i}
    &= \log_{10} \left(\int_{a_i}^1 \frac{da'}{a'^2 H_i(a')} \Big/ \int_{a_i}^1 \frac{da'}{a'^2 H(a')}  \right) \\
    &= \log_{10} \left( \frac{1}{1 + \delta_i} \right) \approx -\frac{\delta_i}{\ln 10} \,.
\end{align}
Similar to the SNe, this distance measurement is also degenerate with $h$. Using $h = 0.7605$ from the DESI analysis, we recalibrate $k_i$ to the Planck cosmology by $k_i \rightarrow k_i + \log_{10}(h_\text{DESI}/h_\text{Planck})$. The recalibration centers the distribution of $h$ around $h_\text{Planck}$.

We consider 4063 galaxies with $0.023 < z < 0.1$ and median $z=0.075$.
We construct $\delta_{\rm FP}$ and $\sigma_{\rm FP}$ maps by inverse-variance averaging the FP galaxy measurements on a HEALPix grid (analogous to \Eq\ref{eqn:inv_var}). The maps and the pixel histograms are shown in \Figs \ref{fig:maps} and \ref{fig:hists}. The fiducial analysis corrects the redshift with the 2M++ peculiar velocity (PV) model \cite{saidJointAnalysis6dFGS2020}, while a systematic variant does not. 
\\\\
{\noindent\it The Cosmic Microwave Background.}
The CMB temperature power spectrum $C_\ell$ is another sensitive probe of $h$ \cite{dodelsonModernCosmology2003, planckcollaborationPlanck2018Results2020a}. To construct the Hubble residual map, we partition the CMB temperature map into 768 HEALPix patches (labeled by $p$), measure the local power spectrum $C_\ell^\text{local}$, and finally estimate the residual $\dcmb(p) = h(p) / h_\text{Planck} - 1$ through a likelihood analysis. 

We use the two Planck 2018 half-mission SMICA maps \cite{planckcollaborationPlanck2018Results2020d, planckcollaborationPlanck2018Results2020a}, $T_1$ and $T_2$ with a resolution of \texttt{nside} = 1024. We use the combined 100, 143, and 217 GHz frequency masks ($f_\mathrm{sky}\approx0.5$) to mitigate galactic foreground and point source contamination. For each pixel $p$, we gnomonically project $T_1$, $T_2$, and the masks to a Cartesian patch. We apodize the mask, measure the mode-decoupled $C_\ell$, and compute the Gaussian covariance of $C_\ell$, $\Sigma$, using \texttt{pymaster} \cite{alonsoUnifiedPseudo$C_ell$Framework2019a}. Since the two half-mission temperature maps measure the same CMB structures but independent noise realizations, their cross-spectrum directly estimates the temperature auto-spectrum without noise bias. 


For each sky patch, we measured the local power spectrum $C_\ell^{\text{obs}}$, and estimate the local $h$ using a maximum-likelihood approach using the two-point statistics. The theoretical model for $C_\ell^{\text{obs}}$ is
\begin{equation}
    C_\ell^{\rm local} = C_\ell^{\rm th} \cdot P_{\ell,1024}^2 \cdot P_{\ell,\text{proj}}^2 \cdot B_\ell^2 \cdot T_{\ell,\text{SMICA}},
\end{equation}
where $C_\ell^\text{th}$ is the full-sky LCDM prediction by \texttt{camb} \cite{lewisEfficientComputationCMB2000}, $P_{1024}$ and $P_{\text{proj}}$ are the pixel window functions of the HEALPix and the Cartesian projection, $B$ and $T_{\text{SMICA}}$ are the SMICA beam and transfer function \cite{planckcollaborationPlanck2018Results2020d}. Since we are interested in the Hubble flow and not the CMB physics, we fix all other parameters (including $\Omega_mh^2, \Omega_bh^2$) and vary only $h$ as we compute the posterior
\begin{equation}
\small
\log \mathcal{P}(h) \propto \sum_{\ell, \ell'} \left(C_\ell^{\text{obs}} - C_\ell^{\rm local}(h)\right) \Sigma^{-1}_{\ell \ell'} \left(C_{\ell'}^{\text{obs}} - C_{\ell'}^{\rm local}(h)\right).
\end{equation}
We construct the residual maps $\dcmb(p) = h(p) / h_{\text{Planck}} - 1$ and $\scmb = \sigma(p) / h_{\text{Planck}}$. The maps and the pixel histograms are shown in \Figs\ref{fig:maps} and \ref{fig:hists}.

In the fiducial analysis, we consider the scale cut $\ell \in [100, 1250]$ in the likelihood analysis. 
A more conservative systematic analysis uses a more aggressive scale-cut $\ell \in [200, 1000]$. Most of the $h$ sensitivity lies near the second acoustic peak at $\ell \approx 500$, which falls within both scale cuts. Larger scales exceed the patch size and are noisy; smaller scales are removed due to potential point source contamination. We also assess the impact of cosmological uncertainty in $\dcmb$ by creating 1000 maps with cosmology sampled from the Planck posterior \cite{planckcollaborationPlanck2018Results2020a}.
\section{Analysis}
We jointly analyze the statistics of the three Hubble residual maps: $\dsn$, $\dfp$, and $\dcmb$. To account for incomplete sky coverage and spatially varying measurement uncertainties, we focus on an inverse-variance weighted version of the $C_\ell$ statistics 
\cite{alonsoUnifiedPseudo$C_ell$Framework2019a,soltisPercentLevelTestIsotropic2019}, 
\begin{align}
    \label{eqn:alm}
    \tilde a_{\ell m}^{\alpha} &= \frac{\int M(\hat {\bf n}) \delta_{\alpha}(\hat{\bf n}) \sigma_{\alpha}^{-2}(\hat{\bf n}) Y_{\ell m}^*(\hat{\bf n}) \, d\Omega(\hat{\bf n})}{\int M(\hat {\bf n}) \sigma_{\alpha}^{-2}(\hat{\bf n}) \, d\Omega(\hat{\bf n}) / 4 \pi} \,, \\
    \label{eqn:cl}
    \tilde C_\ell^{\alpha,\beta} &= \langle \tilde a_{\ell m}^{\alpha} \tilde a_{\ell m}^{\beta,*} \rangle_m \,,
\end{align}
where $\ell \in [1, 23]$, $\alpha,\beta$ refer to SNe, FP, or CMB, and $M(\hat {\bf n})$ is the mask. \Eqs\ref{eqn:alm} and \ref{eqn:cl} are closely related to the standard $C_\ell$ statistics -- if the unmasked data have spatially constant uncertainty and a flat power spectrum, \Eq\ref{eqn:alm} reduces to 
$
\tilde a_{\ell m} = \fsky^{-1}
    \int M(\hat{\bf n})\delta(\hat{\bf n}) Y_{\ell m}^*(\hat{\bf n}) \, d\Omega(\hat{\bf n}),
$
and the weighted power spectrum is related to the standard one by
\begin{equation}
    C_\ell \approx \fsky \tilde C_\ell \,.
\end{equation}

To test for isotropy, we extend the approach of \citet{soltisPercentLevelTestIsotropic2019} to multiple tracers. Consider the set $\{(\delta(p), \sigma(p))\}$ for all \emph{unmasked} pixels $p$. This \emph{distribution} includes contributions from both the potentially anisotropic Hubble field and the underlying noise properties. For example, it captures how the spatially varying mask fraction of the CMB affects $\dcmb$ and $\scmb$, and how individual object measurement uncertainties and spatial number density variations affect the SNe and FP maps. 
For each $\delta$ map, we create 1000 \emph{isotropic noise realizations} by drawing unmasked pixel values from this distribution $\{(\delta(p), \sigma(p))\}$. For each realization, we compute the auto- and cross-power spectra $\tilde C_\ell^{{\rm iso},\alpha\beta}$ among all tracers. 
\Fig\ref{fig:cl_cross} compares $\tilde C_\ell$ and $\tilde C_\ell^{{\rm iso},\alpha\beta}$ for the fiducial (black) and the systematics models (blue).

To quantitatively evaluate how well each tracer individually, as well as their combination, supports or rejects the isotropy hypothesis, we perform a $\chi^2$ test on the auto- and cross-spectra. We consider the multipole range $\ell \in [1, 23]$ and estimate the covariance matrix from the isotropic noise realizations. To account for finite sample bias in the inverse covariance estimator, we apply the correction \cite{hartlapNonGaussianityCosmicShear2009, dodelsonEffectCovarianceEstimator2013}
\begin{equation}
\chi^2 = \chi^2_{\rm raw} \times \frac{N_{\rm sim} - N_{\rm data} - 2}{N_{\rm sim} - 1} \,,
\end{equation}
where $N_{\rm sim}$ is the number of isotropic simulations and $N_{\rm data}$ is the length of the data vector.

\begin{table}[htbp]
\centering
\begin{tabular*}{\linewidth}{@{\extracolsep{\fill}} l c l cc}
\toprule
\multicolumn{3}{c}{Configuration} & $\chi^2$/DoF & $p$-value \\
\midrule
\multirow{4}{*}{\begin{tabular}{@{}l@{}}SN + CMB\\\small\textit{(DoF=66)}\end{tabular}}
 & & \textbf{Fiducial}            & \textbf{0.92} & \textbf{0.66} \\
\cmidrule{2-5}
 & \multirow{3}{*}{\rotatebox{90}{\small sys. variants}}
 & SN: $z_{\rm CMB}$   & 1.19 & 0.14          \\
 & & CMB: scale-cut      & 0.96 & 0.57          \\
 & & CMB: cosmo. uncert. & 0.88 & 0.75          \\
\cmidrule{1-1}\cmidrule{3-5}
\multirow{2}{*}{\begin{tabular}{@{}l@{}}SN+FP+CMB\\\small\textit{(DoF=132)}\end{tabular}}
 & & Fiducial & 1.02 & 0.41 \\
 & & FP: no PV & 1.21  & 0.05 \\
\bottomrule
\end{tabular*}
\caption{Results of the isotropy $\chi^2$ test using SNe, CMB, and FP tracers. The main result uses only the SNe and the CMB data (row 1), and supports the null hypothesis of isotropy. Lower rows show the test statistics when varying a single tracer's systematic model or combining with the FP data.}
\label{tab:isotropy_sn_cmb}
\end{table}
\Fig\ref{fig:cl_cross} shows the test statistics of the individual auto- and cross-correlations. The auto and cross-spectra of the fiducial $\dsn$ and $\dcmb$ maps are consistent with isotropy. 
Their combined statistics have a p-value of 0.66 (\Tab\ref{tab:isotropy_sn_cmb}, row 1) and supports the null hypotheses of \emph{(1) isotropic expansion and (2) uncorrelated expansion across epochs.} This is the main finding of this letter. The result only holds after SNe peculiar velocity correction, and is robust against conservative CMB scale cuts ($p=0.57$) and cosmology uncertainty marginalization ($p=0.75$).

For $\dfp$, we find a significant excess of power around $\ell = 6$ ($\theta\approx30^\circ$, $p=0.07$). Unlike the SNe, the PV correction only alleviates, but cannot eliminate, this anomaly. However, since there is no significant cross-correlation between $\dfp$ and $\dsn$ (which cover the same low-redshift volume) and $\dcmb$, we are reluctant to associate the FP anomaly with true cosmic anisotropy. Another reason against this association is that we find a significant correlation between $\dfp$ and the FP number density distribution (and to a lesser degree the mean redshift in each pixel) in the same $\ell$ range. This evidence for a possible systematic effect in the FP sample is robust across all aforementioned systematic choices. The current FP sample comprises only targets from the DESI science validation phase and has significant spatial depth variation across the sky, potentially inducing unaccounted-for selection effects. Future DESI data expects a more homogeneous coverage \cite{saulderTargetSelectionDESI2023, saidDESIPeculiarVelocity2024}.

\begin{figure}
    \centering
    \includegraphics[width=1.0\linewidth]{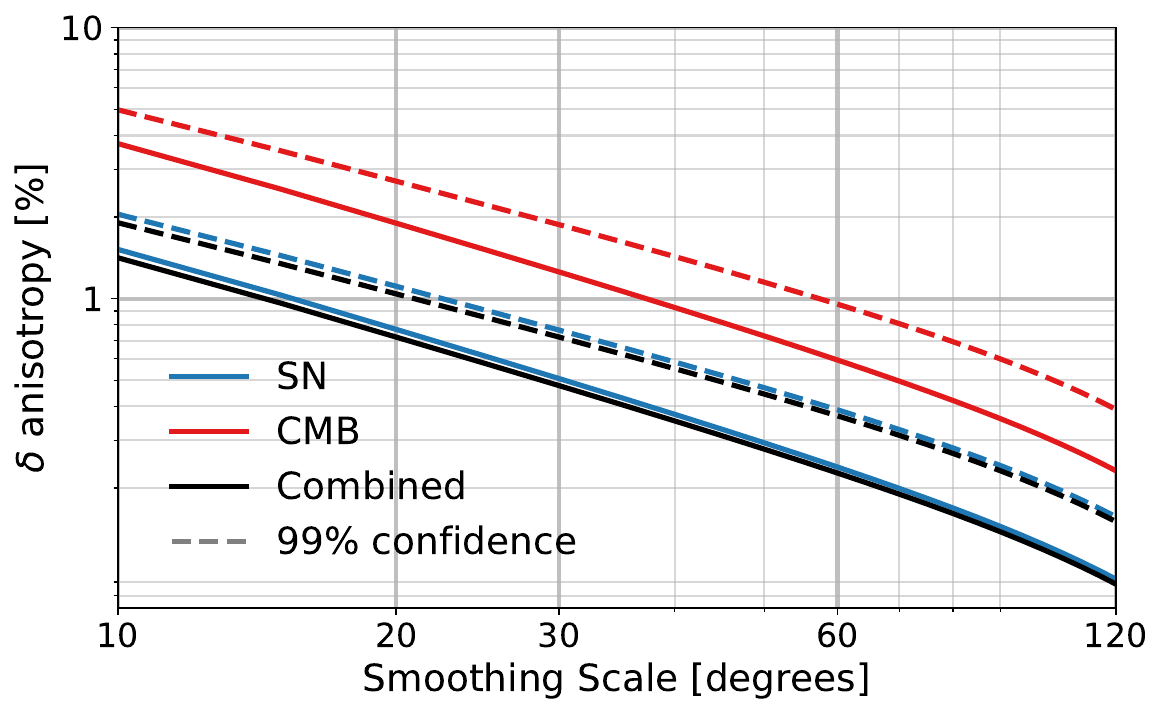}
        \caption{Upper limits on Hubble constant anisotropy as a function of smoothing scale. Solid lines show the mean noise level from isotropic realizations, while dashed lines indicate 99\% confidence upper limits for the SNe, CMB, and the combined maps
        }
    \label{fig:limits}
\end{figure}

To obtain an upper limit on the variance of the anisotropy, we follow the general approach of \citet{soltisPercentLevelTestIsotropic2019} and decompose the observed $\delta$ into a true signal and noise, $\dobs = \dtrue + \noise$. Since $\dobs$ for both SNe and CMB are consistent with isotropy, the noise must dominate $\dobs$, thereby constraining the amplitude of $\dtrue$. We construct this limit using both probes via the full-sky power spectrum. Since $\dobs \approx \noise$, the 99\% upper bound on the variance of the true $\delta$
\begin{equation}
    \left\langle \dtrue^{2} \right\rangle_{99\%}
    \;<\;
    \sum_{\ell} \frac{2\ell+1}{4\pi}
    \Bigl(C_\ell^{\rm iso}
          + 2.33\,\Delta C_\ell^{\rm iso}\Bigr)
    \,B_\ell^{2}\,d\ell \,,
\end{equation}
where $C_\ell^{\rm iso}$ and $\Delta C_\ell^{\rm iso}$ are the mean and scatter of the power spectrum of the full-sky isotropic noise, and $B_\ell$ smooths the map via a Gaussian beam. The factor 2.33 corresponds to the one-sided 99\% confidence threshold for a Gaussian distribution. We derive this limit for the SNe, CMB, and inverse-variance combined maps on their joint footprint for scales between 10 and 120 degrees in \Fig\ref{fig:limits}. At 60 degrees, the 99\% confidence limits on low-$z$ and high-$z$ anisotropy are 0.39\% and 0.95\% respectively, and 0.37\% when combined. 

\section{Conclusion}
We tested the isotropy of cosmic expansion using three independent distance tracers: Type Ia supernovae, fundamental plane galaxies, and CMB temperature fluctuations. Aside from a localized anomaly in the FP auto-power spectrum around $\ell = 6$, all measurements are consistent with isotropic expansion at low and high redshift, and results are robust across the systematic modeling choices we tested. The SNe and CMB measurements yield stringent constraints on potential Hubble flow anisotropy, with 99\% confidence upper limits of 0.39\% for low-redshift SNe, 0.95\% for high-redshift CMB, and 0.37\% when combined at a 60-degree smoothing scale. 

We argue that -- given this other evidence -- it is premature to claim that the FP anomaly is a sign of genuine cosmic anisotropy. This hesitance is supported by the absence of significant cross-correlation between $\dfp$ and the other tracers, particularly with $\dsn$ which probes the same low-redshift volume, and by the correlation between $\dfp$ and the spatial distribution of FP galaxy density. Future DESI data releases with more homogeneous sky coverage should shed light on this anomaly.

Our results agree with and extend those of Refs.~\cite{soltisPercentLevelTestIsotropic2019,kimLimitsAnisotropicInflation2013}: multiple probes establish that any small Hubble residuals are uncorrelated over low and high redshifts. Our results are complimentary to those of Refs.~\cite{Cowell_2022,Hu2024_CPtest}, which focus on low-redshift anisotropy on the largest scales with particular shapes: we jointly studied both early and late-Universe anisotropy, and only focused on the power spectrum.

{\noindent\it Acknowledgments.} 
AZ thanks Rachel Mandelbaum for early discussions and Yuuki Omori for help with the Planck data. AZ is supported by the Jane Street Graduate Research Fellowship. This work was supported by FermiForward Discovery Group, LLC, under Contract No. 89243024CSC000002 with the U.S. Department of Energy, Office of Science, Office of High Energy Physics. DS acknowledges support from Department of Energy grant DE-SC0010007, the David and Lucile Packard Foundation, and the Templeton Foundation grant focused on understanding motions in the nearby universe.

\bibliography{ref_alan, ref_rev}

@Article{Hu2024_CPtest,
  author    = {Hu, Jian-Ping and Wang, Yu-Yan and Hu, Jun and Wang, Fei-Yu},
  title     = {Testing the cosmological principle with the Pantheon+ sample and the region-fitting method},
  journal   = {Astronomy \& Astrophysics},
  year      = {2024},
  volume    = {681},
  pages     = {A88},
  doi       = {10.1051/0004-6361/202347121},
  eprint    = {2310.11727},
  eprinttype= {arXiv}
}

@article{alonsoUnifiedPseudo$C_ell$Framework2019a,
  title = {A Unified Pseudo-\${{C}}\_{\textbackslash}ell\$ Framework},
  author = {Alonso, David and Sanchez, Javier and Slosar, An{\v z}e},
  year = {2019},
  month = apr,
  journal = {Monthly Notices of the Royal Astronomical Society},
  volume = {484},
  number = {3},
  eprint = {1809.09603},
  primaryclass = {astro-ph},
  pages = {4127--4151},
  issn = {0035-8711, 1365-2966},
  doi = {10.1093/mnras/stz093},
  urldate = {2024-12-24},
  archiveprefix = {arXiv}
}

@article{bennettSevenYearWilkinsonMicrowave2011,
  title = {Seven-{{Year Wilkinson Microwave Anisotropy Probe}} ({{WMAP}}) {{Observations}}: {{Are There Cosmic Microwave Background Anomalies}}?},
  shorttitle = {Seven-{{Year Wilkinson Microwave Anisotropy Probe}} ({{WMAP}}) {{Observations}}},
  author = {Bennett, C. L. and Hill, R. S. and Hinshaw, G. and Larson, D. and Smith, K. M. and Dunkley, J. and Gold, B. and Halpern, M. and Jarosik, N. and Kogut, A. and Komatsu, E. and Limon, M. and Meyer, S. S. and Nolta, M. R. and Odegard, N. and Page, L. and Spergel, D. N. and Tucker, G. S. and Weiland, J. L. and Wollack, E. and Wright, E. L.},
  year = {2011},
  month = feb,
  journal = {The Astrophysical Journal Supplement Series},
  volume = {192},
  number = {2},
  eprint = {1001.4758},
  primaryclass = {astro-ph},
  pages = {17},
  issn = {0067-0049, 1538-4365},
  doi = {10.1088/0067-0049/192/2/17},
  urldate = {2024-08-07},
  archiveprefix = {arXiv}
}

@article{colinProbingAnisotropicLocal2011,
  title = {Probing the Anisotropic Local Universe and beyond with {{SNe Ia}} Data},
  author = {Colin, Jacques and Mohayaee, Roya and Sarkar, Subir and Shafieloo, Arman},
  year = {2011},
  month = jun,
  journal = {Monthly Notices of the Royal Astronomical Society},
  volume = {414},
  number = {1},
  eprint = {1011.6292},
  primaryclass = {astro-ph},
  pages = {264--271},
  issn = {00358711},
  doi = {10.1111/j.1365-2966.2011.18402.x},
  urldate = {2025-05-24},
  archiveprefix = {arXiv}
}

@article{collaborationDESI2024VI2025,
  title = {{{DESI}} 2024 {{VI}}: {{Cosmological Constraints}} from the {{Measurements}} of {{Baryon Acoustic Oscillations}}},
  shorttitle = {{{DESI}} 2024 {{VI}}},
  author = {Collaboration, {\relax DESI} and Adame, A. G. and Aguilar, J. and Ahlen, S. and Alam, S. and Alexander, D. M. and Alvarez, M. and Alves, O. and Anand, A. and Andrade, U. and Armengaud, E. and Avila, S. and Aviles, A. and Awan, H. and {Bahr-Kalus}, B. and Bailey, S. and Baltay, C. and Bault, A. and Behera, J. and BenZvi, S. and Bera, A. and Beutler, F. and Bianchi, D. and Blake, C. and Blum, R. and Brieden, S. and Brodzeller, A. and Brooks, D. and {Buckley-Geer}, E. and Burtin, E. and Calderon, R. and Canning, R. and Rosell, A. Carnero and Cereskaite, R. and {Cervantes-Cota}, J. L. and Chabanier, S. and Chaussidon, E. and {Chaves-Montero}, J. and Chen, S. and Chen, X. and Claybaugh, T. and Cole, S. and Cuceu, A. and Davis, T. M. and Dawson, K. and de la Macorra, A. and de Mattia, A. and Deiosso, N. and Dey, A. and Dey, B. and Ding, Z. and Doel, P. and Edelstein, J. and Eftekharzadeh, S. and Eisenstein, D. J. and Elliott, A. and Fagrelius, P. and Fanning, K. and Ferraro, S. and Ereza, J. and Findlay, N. and Flaugher, B. and {Font-Ribera}, A. and {Forero-S{\'a}nchez}, D. and {Forero-Romero}, J. E. and Frenk, C. S. and {Garcia-Quintero}, C. and Gazta{\~n}aga, E. and {Gil-Mar{\'i}n}, H. and Gontcho, S. Gontcho A. and {Gonzalez-Morales}, A. X. and {Gonzalez-Perez}, V. and Gordon, C. and Green, D. and Gruen, D. and Gsponer, R. and Gutierrez, G. and Guy, J. and Hadzhiyska, B. and Hahn, C. and Hanif, M. M. S. and {Herrera-Alcantar}, H. K. and Honscheid, K. and Howlett, C. and Huterer, D. and Ir{\v s}i{\v c}, V. and Ishak, M. and Juneau, S. and Kara{\c c}ayl{\i}, N. G. and Kehoe, R. and Kent, S. and Kirkby, D. and Kremin, A. and Krolewski, A. and Lai, Y. and Lan, T.-W. and Landriau, M. and Lang, D. and Lasker, J. and Goff, J. M. Le and Guillou, L. Le and Leauthaud, A. and Levi, M. E. and Li, T. S. and Linder, E. and Lodha, K. and Magneville, C. and Manera, M. and Margala, D. and Martini, P. and Maus, M. and McDonald, P. and {Medina-Varela}, L. and Meisner, A. and {Mena-Fern{\'a}ndez}, J. and Miquel, R. and Moon, J. and Moore, S. and Moustakas, J. and Mudur, N. and Mueller, E. and {Mu{\~n}oz-Guti{\'e}rrez}, A. and Myers, A. D. and Nadathur, S. and Napolitano, L. and Neveux, R. and Newman, J. A. and Nguyen, N. M. and Nie, J. and Niz, G. and Noriega, H. E. and Padmanabhan, N. and Paillas, E. and {Palanque-Delabrouille}, N. and Pan, J. and Penmetsa, S. and Percival, W. J. and Pieri, M. M. and Pinon, M. and Poppett, C. and Porredon, A. and Prada, F. and {P{\'e}rez-Fern{\'a}ndez}, A. and {P{\'e}rez-R{\`a}fols}, I. and Rabinowitz, D. and Raichoor, A. and {Ram{\'i}rez-P{\'e}rez}, C. and {Ramirez-Solano}, S. and Ravoux, C. and Rashkovetskyi, M. and Rezaie, M. and Rich, J. and Rocher, A. and Rockosi, C. and Roe, N. A. and {Rosado-Marin}, A. and Ross, A. J. and Rossi, G. and Ruggeri, R. and {Ruhlmann-Kleider}, V. and Samushia, L. and Sanchez, E. and Saulder, C. and Schlafly, E. F. and Schlegel, D. and Schubnell, M. and Seo, H. and Shafieloo, A. and Sharples, R. and Silber, J. and Slosar, A. and Smith, A. and Sprayberry, D. and Tan, T. and Tarl{\'e}, G. and Taylor, P. and Trusov, S. and {Ure{\~n}a-L{\'o}pez}, L. A. and Vaisakh, R. and Valcin, D. and Valdes, F. and {Vargas-Maga{\~n}a}, M. and Verde, L. and Walther, M. and Wang, B. and Wang, M. S. and Weaver, B. A. and Weaverdyck, N. and Wechsler, R. H. and Weinberg, D. H. and White, M. and Yu, J. and Yu, Y. and Yuan, S. and Y{\`e}che, C. and Zaborowski, E. A. and Zarrouk, P. and Zhang, H. and Zhao, C. and Zhao, R. and Zhou, R. and Zhuang, T. and Zou, H.},
  year = {2025},
  month = feb,
  journal = {Journal of Cosmology and Astroparticle Physics},
  volume = {2025},
  number = {02},
  eprint = {2404.03002},
  primaryclass = {astro-ph},
  pages = {021},
  issn = {1475-7516},
  doi = {10.1088/1475-7516/2025/02/021},
  urldate = {2025-06-08},
  archiveprefix = {arXiv}
}

@misc{collaborationDESIDR2Results2025,
  title = {{{DESI DR2 Results II}}: {{Measurements}} of {{Baryon Acoustic Oscillations}} and {{Cosmological Constraints}}},
  shorttitle = {{{DESI DR2 Results II}}},
  author = {Collaboration, {\relax DESI} and {Abdul-Karim}, M. and Aguilar, J. and Ahlen, S. and Alam, S. and Allen, L. and Prieto, C. Allende and Alves, O. and Anand, A. and Andrade, U. and Armengaud, E. and Aviles, A. and Bailey, S. and Baltay, C. and Bansal, P. and Bault, A. and Behera, J. and BenZvi, S. and Bianchi, D. and Blake, C. and Brieden, S. and Brodzeller, A. and Brooks, D. and {Buckley-Geer}, E. and Burtin, E. and Calderon, R. and Canning, R. and Rosell, A. Carnero and Carrilho, P. and Casas, L. and Castander, F. J. and Cereskaite, R. and Charles, M. and Chaussidon, E. and {Chaves-Montero}, J. and Chebat, D. and Chen, X. and Claybaugh, T. and Cole, S. and Cooper, A. P. and Cuceu, A. and Dawson, K. S. and de la Macorra, A. and de Mattia, A. and Deiosso, N. and Costa, J. Della and Demina, R. and Dey, A. and Dey, B. and Ding, Z. and Doel, P. and Edelstein, J. and Eisenstein, D. J. and Elbers, W. and Fagrelius, P. and Fanning, K. and {Fern{\'a}ndez-Garc{\'i}a}, E. and Ferraro, S. and {Font-Ribera}, A. and {Forero-Romero}, J. E. and Frenk, C. S. and {Garcia-Quintero}, C. and Garrison, L. H. and Gazta{\~n}aga, E. and {Gil-Mar{\'i}n}, H. and Gontcho, S. Gontcho A. and Gonzalez, D. and {Gonzalez-Morales}, A. X. and Gordon, C. and Green, D. and Gutierrez, G. and Guy, J. and Hadzhiyska, B. and Hahn, C. and He, S. and Herbold, M. and {Herrera-Alcantar}, H. K. and Ho, M. and Honscheid, K. and Howlett, C. and Huterer, D. and Ishak, M. and Juneau, S. and Kamble, N. V. and Kara{\c c}ayl{\i}, N. G. and Kehoe, R. and Kent, S. and Kim, A. G. and Kirkby, D. and Kisner, T. and Koposov, S. E. and Kremin, A. and Krolewski, A. and Lahav, O. and Lamman, C. and Landriau, M. and Lang, D. and Lasker, J. and Goff, J. M. Le and Guillou, L. Le and Leauthaud, A. and Levi, M. E. and Li, Q. and Li, T. S. and Lodha, K. and Lokken, M. and {Lozano-Rodr{\'i}guez}, F. and Magneville, C. and Manera, M. and Martini, P. and Matthewson, W. L. and Meisner, A. and {Mena-Fern{\'a}ndez}, J. and Menegas, A. and Mergulh{\~a}o, T. and Miquel, R. and Moustakas, J. and {Mu{\~n}oz-Guti{\'e}rrez}, A. and {Mu{\~n}oz-Santos}, D. and Myers, A. D. and Nadathur, S. and Naidoo, K. and Napolitano, L. and Newman, J. A. and Niz, G. and Noriega, H. E. and Paillas, E. and {Palanque-Delabrouille}, N. and Pan, J. and Peacock, J. and Ibanez, Marcos Pellejero and Percival, W. J. and {P{\'e}rez-Fern{\'a}ndez}, A. and {P{\'e}rez-R{\`a}fols}, I. and Pieri, M. M. and Poppett, C. and Prada, F. and Rabinowitz, D. and Raichoor, A. and {Ram{\'i}rez-P{\'e}rez}, C. and Rashkovetskyi, M. and Ravoux, C. and Rich, J. and Rocher, A. and Rockosi, C. and Rohlf, J. and {Rom{\'a}n-Herrera}, J. O. and Ross, A. J. and Rossi, G. and Ruggeri, R. and {Ruhlmann-Kleider}, V. and Samushia, L. and Sanchez, E. and Sanders, N. and Schlegel, D. and Schubnell, M. and Seo, H. and Shafieloo, A. and Sharples, R. and Silber, J. and Sinigaglia, F. and Sprayberry, D. and Tan, T. and Tarl{\'e}, G. and Taylor, P. and Turner, W. and {Ure{\~n}a-L{\'o}pez}, L. A. and Vaisakh, R. and Valdes, F. and Valogiannis, G. and {Vargas-Maga{\~n}a}, M. and Verde, L. and Walther, M. and Weaver, B. A. and Weinberg, D. H. and White, M. and Wolfson, M. and Y{\`e}che, C. and Yu, J. and Zaborowski, E. A. and Zarrouk, P. and Zhai, Z. and Zhang, H. and Zhao, C. and Zhao, G. B. and Zhou, R. and Zou, H.},
  year = {2025},
  month = mar,
  number = {arXiv:2503.14738},
  eprint = {2503.14738},
  primaryclass = {astro-ph},
  publisher = {arXiv},
  doi = {10.48550/arXiv.2503.14738},
  urldate = {2025-06-09},
  archiveprefix = {arXiv}
}

@article{coorayMeasuringDarkEnergy2010,
  title = {Measuring Dark Energy Spatial Inhomogeneity with Supernova Data},
  author = {Cooray, Asantha and Holz, Daniel E. and Caldwell, Robert},
  year = {2010},
  month = nov,
  journal = {Journal of Cosmology and Astroparticle Physics},
  volume = {2010},
  number = {11},
  eprint = {0812.0376},
  primaryclass = {astro-ph},
  pages = {015--015},
  issn = {1475-7516},
  doi = {10.1088/1475-7516/2010/11/015},
  urldate = {2025-06-11},
  archiveprefix = {arXiv}
}

@article{copelandDynamicsDarkEnergy2006,
  title = {Dynamics of Dark Energy},
  author = {Copeland, Edmund J. and Sami, M. and Tsujikawa, Shinji},
  year = {2006},
  month = nov,
  journal = {International Journal of Modern Physics D},
  volume = {15},
  number = {11},
  eprint = {hep-th/0603057},
  pages = {1753--1935},
  issn = {0218-2718, 1793-6594},
  doi = {10.1142/S021827180600942X},
  urldate = {2025-06-09},
  archiveprefix = {arXiv}
}

@article{copiLargeangleAnomaliesCMB2010,
  title = {Large-Angle Anomalies in the {{CMB}}},
  author = {Copi, Craig J. and Huterer, Dragan and Schwarz, Dominik J. and Starkman, Glenn D.},
  year = {2010},
  month = jan,
  journal = {Advances in Astronomy},
  volume = {2010},
  number = {1},
  eprint = {1004.5602},
  primaryclass = {astro-ph, physics:gr-qc, physics:hep-th},
  pages = {847541},
  issn = {1687-7969, 1687-7977},
  doi = {10.1155/2010/847541},
  urldate = {2024-08-07},
  archiveprefix = {arXiv}
}

@article{deoliveira-costaSignificanceLargestScale2004,
  title = {Significance of the Largest Scale {{CMB}} Fluctuations in {{WMAP}}},
  author = {{De Oliveira-Costa}, Ang{\'e}lica and Tegmark, Max and Zaldarriaga, Matias and Hamilton, Andrew},
  year = {2004},
  month = mar,
  journal = {Physical Review D},
  volume = {69},
  number = {6},
  pages = {063516},
  issn = {1550-7998, 1550-2368},
  doi = {10.1103/PhysRevD.69.063516},
  urldate = {2025-05-24},
  copyright = {http://link.aps.org/licenses/aps-default-license}
}

@article{dodelsonEffectCovarianceEstimator2013,
  title = {The {{Effect}} of {{Covariance Estimator Error}} on {{Cosmological Parameter Constraints}}},
  author = {Dodelson, Scott and Schneider, Michael D.},
  year = {2013},
  month = sep,
  journal = {Physical Review D},
  volume = {88},
  number = {6},
  eprint = {1304.2593},
  primaryclass = {astro-ph},
  pages = {063537},
  issn = {1550-7998, 1550-2368},
  doi = {10.1103/PhysRevD.88.063537},
  urldate = {2025-05-29},
  archiveprefix = {arXiv}
}

@book{dodelsonModernCosmology2003,
  title = {Modern Cosmology},
  author = {Dodelson, Scott},
  year = {2003},
  publisher = {Academic Press},
  address = {San Diego, Calif},
  isbn = {978-0-12-219141-1},
  lccn = {QB981 .D63 2003}
}

@article{eriksenPointCorrelationFunctions2005,
  title = {The {{{\emph{N}}}} -{{Point Correlation Functions}} of the {{First}}-{{Year}} {{{\emph{Wilkinson Microwave Anisotropy Probe}}}} {{Sky Maps}}},
  author = {Eriksen, H. K. and Banday, A. J. and Gorski, K. M. and Lilje, P. B.},
  year = {2005},
  month = mar,
  journal = {The Astrophysical Journal},
  volume = {622},
  number = {1},
  pages = {58--71},
  issn = {0004-637X, 1538-4357},
  doi = {10.1086/427897},
  urldate = {2025-06-09}
}

@article{fosalbaExplainingCosmologicalAnisotropy2021,
  title = {Explaining {{Cosmological Anisotropy}}: {{Evidence}} for {{Causal Horizons}} from {{CMB}} Data},
  shorttitle = {Explaining {{Cosmological Anisotropy}}},
  author = {Fosalba, Pablo and Gaztanaga, Enrique},
  year = {2021},
  month = may,
  journal = {Monthly Notices of the Royal Astronomical Society},
  volume = {504},
  number = {4},
  eprint = {2011.00910},
  primaryclass = {astro-ph},
  pages = {5840--5862},
  issn = {0035-8711, 1365-2966},
  doi = {10.1093/mnras/stab1193},
  urldate = {2025-05-24},
  archiveprefix = {arXiv}
}

@misc{gimeno-amoExploringStatisticalIsotropy2025,
  title = {Exploring {{Statistical Isotropy}} in {{Planck Data Release}} 4: {{Angular Clustering}} and {{Cosmological Parameter Variations Across}} the {{Sky}}},
  shorttitle = {Exploring {{Statistical Isotropy}} in {{Planck Data Release}} 4},
  author = {{Gimeno-Amo}, C. and Hansen, F. K. and {Mart{\'i}nez-Gonz{\'a}lez}, E. and Barreiro, R. B. and Banday, A. J.},
  year = {2025},
  month = apr,
  number = {arXiv:2504.05597},
  eprint = {2504.05597},
  primaryclass = {astro-ph},
  publisher = {arXiv},
  doi = {10.48550/arXiv.2504.05597},
  urldate = {2025-06-12},
  archiveprefix = {arXiv}
}

@article{gorskiHEALPixFrameworkHigh2005,
  title = {{{HEALPix}} -- a {{Framework}} for {{High Resolution Discretization}}, and {{Fast Analysis}} of {{Data Distributed}} on the {{Sphere}}},
  author = {Gorski, K. M. and Hivon, E. and Banday, A. J. and Wandelt, B. D. and Hansen, F. K. and Reinecke, M. and Bartelman, M.},
  year = {2005},
  month = apr,
  journal = {The Astrophysical Journal},
  volume = {622},
  number = {2},
  eprint = {astro-ph/0409513},
  pages = {759--771},
  issn = {0004-637X, 1538-4357},
  doi = {10.1086/427976},
  urldate = {2022-12-09},
  archiveprefix = {arXiv}
}

@misc{haridasuRadialTullyFisherRelation2024,
  title = {Radial {{Tully-Fisher}} Relation and the Local Variance of {{Hubble}} Parameter},
  author = {Haridasu, Balakrishna S. and Salucci, Paolo and Sharma, Gauri},
  year = {2024},
  month = mar,
  number = {arXiv:2403.06859},
  eprint = {2403.06859},
  primaryclass = {astro-ph},
  publisher = {arXiv},
  doi = {10.48550/arXiv.2403.06859},
  urldate = {2025-05-24},
  archiveprefix = {arXiv}
}

@article{hartlapNonGaussianityCosmicShear2009,
  title = {The Non-{{Gaussianity}} of the Cosmic Shear Likelihood - or: {{How}} Odd Is the {{Chandra Deep Field South}}?},
  shorttitle = {The Non-{{Gaussianity}} of the Cosmic Shear Likelihood - Or},
  author = {Hartlap, J. and Schrabback, T. and Simon, P. and Schneider, P.},
  year = {2009},
  month = sep,
  journal = {Astronomy \& Astrophysics},
  volume = {504},
  number = {3},
  eprint = {0901.3269},
  primaryclass = {astro-ph},
  pages = {689--703},
  issn = {0004-6361, 1432-0746},
  doi = {10.1051/0004-6361/200911697},
  urldate = {2025-05-29},
  archiveprefix = {arXiv}
}

@article{hinshawFirstYearWilkinson2003,
  title = {First {{Year Wilkinson Microwave Anisotropy Probe}} ({{WMAP}}) {{Observations}}: {{Angular Power Spectrum}}},
  shorttitle = {First {{Year Wilkinson Microwave Anisotropy Probe}} ({{WMAP}}) {{Observations}}},
  author = {Hinshaw, G. and Spergel, D. N. and Verde, L. and Hill, R. S. and Meyer, S. S. and Barnes, C. and Bennett, C. L. and Halpern, M. and Jarosik, N. and Kogut, A. and Komatsu, E. and Limon, M. and Page, L. and Tucker, G. S. and Weiland, J. and Wollack, E. and Wright, E. L.},
  year = {2003},
  month = sep,
  journal = {The Astrophysical Journal Supplement Series},
  volume = {148},
  number = {1},
  eprint = {astro-ph/0302217},
  pages = {135--159},
  issn = {0067-0049, 1538-4365},
  doi = {10.1086/377225},
  urldate = {2025-06-08},
  archiveprefix = {arXiv}
}

@article{jensenInfraredSurfaceBrightness2021,
  title = {Infrared {{Surface Brightness Fluctuation Distances}} for {{MASSIVE}} and {{Type Ia Supernova Host Galaxies}}*},
  author = {Jensen, Joseph B. and Blakeslee, John P. and Ma, Chung-Pei and Milne, Peter A. and Brown, Peter J. and Cantiello, Michele and Garnavich, Peter M. and Greene, Jenny E. and Lucey, John R. and Phan, Anh and Tully, R. Brent and Wood, Charlotte M.},
  year = {2021},
  month = aug,
  journal = {The Astrophysical Journal Supplement Series},
  volume = {255},
  number = {2},
  pages = {21},
  issn = {0067-0049, 1538-4365},
  doi = {10.3847/1538-4365/ac01e7},
  urldate = {2025-05-27}
}

@misc{kamionkowskiHubbleTensionEarly2022,
  title = {The {{Hubble Tension}} and {{Early Dark Energy}}},
  author = {Kamionkowski, Marc and Riess, Adam G.},
  year = {2022},
  month = nov,
  number = {arXiv:2211.04492},
  eprint = {2211.04492},
  primaryclass = {astro-ph},
  publisher = {arXiv},
  doi = {10.48550/arXiv.2211.04492},
  urldate = {2025-06-08},
  archiveprefix = {arXiv}
}

@article{kimLimitsAnisotropicInflation2013,
  title = {Limits on Anisotropic Inflation from the {{Planck}} Data},
  author = {Kim, Jaiseung and Komatsu, Eiichiro},
  year = {2013},
  month = nov,
  journal = {Physical Review D},
  volume = {88},
  number = {10},
  eprint = {1310.1605},
  primaryclass = {astro-ph},
  pages = {101301},
  issn = {1550-7998, 1550-2368},
  doi = {10.1103/PhysRevD.88.101301},
  urldate = {2024-08-07},
  archiveprefix = {arXiv}
}

@article{landAxisEvil2005,
  title = {The Axis of Evil},
  author = {Land, Kate and Magueijo, Joao},
  year = {2005},
  month = aug,
  journal = {Physical Review Letters},
  volume = {95},
  number = {7},
  eprint = {astro-ph/0502237},
  pages = {071301},
  issn = {0031-9007, 1079-7114},
  doi = {10.1103/PhysRevLett.95.071301},
  urldate = {2025-05-24},
  archiveprefix = {arXiv}
}

@article{lewisEfficientComputationCMB2000,
  title = {Efficient {{Computation}} of {{CMB}} Anisotropies in Closed {{FRW}} Models},
  author = {Lewis, Antony and Challinor, Anthony and Lasenby, Anthony},
  year = {2000},
  month = aug,
  journal = {The Astrophysical Journal},
  volume = {538},
  number = {2},
  eprint = {astro-ph/9911177},
  pages = {473--476},
  issn = {0004-637X, 1538-4357},
  doi = {10.1086/309179},
  urldate = {2023-11-25},
  archiveprefix = {arXiv}
}

@article{maleknejadGaugeFieldsInflation2013,
  title = {Gauge {{Fields}} and {{Inflation}}},
  author = {Maleknejad, A. and {Sheikh-Jabbari}, M. M. and Soda, J.},
  year = {2013},
  month = jul,
  journal = {Physics Reports},
  volume = {528},
  number = {4},
  eprint = {1212.2921},
  primaryclass = {hep-th},
  pages = {161--261},
  issn = {03701573},
  doi = {10.1016/j.physrep.2013.03.003},
  urldate = {2025-06-08},
  archiveprefix = {arXiv}
}

@article{planckcollaborationPlanck2018Results2020a,
  title = {Planck 2018 Results. {{VI}}. {{Cosmological}} Parameters},
  author = {Planck Collaboration and Aghanim, N. and Akrami, Y. and Ashdown, M. and Aumont, J. and Baccigalupi, C. and Ballardini, M. and Banday, A. J. and Barreiro, R. B. and Bartolo, N. and Basak, S. and Battye, R. and Benabed, K. and Bernard, J.-P. and Bersanelli, M. and Bielewicz, P. and Bock, J. J. and Bond, J. R. and Borrill, J. and Bouchet, F. R. and Boulanger, F. and Bucher, M. and Burigana, C. and Butler, R. C. and Calabrese, E. and Cardoso, J.-F. and Carron, J. and Challinor, A. and Chiang, H. C. and Chluba, J. and Colombo, L. P. L. and Combet, C. and Contreras, D. and Crill, B. P. and Cuttaia, F. and {de Bernardis}, P. and {de Zotti}, G. and Delabrouille, J. and Delouis, J.-M. and Di Valentino, E. and Diego, J. M. and Dor{\'e}, O. and Douspis, M. and Ducout, A. and Dupac, X. and Dusini, S. and Efstathiou, G. and Elsner, F. and En{\ss}lin, T. A. and Eriksen, H. K. and Fantaye, Y. and Farhang, M. and Fergusson, J. and {Fernandez-Cobos}, R. and Finelli, F. and Forastieri, F. and Frailis, M. and Fraisse, A. A. and Franceschi, E. and Frolov, A. and Galeotta, S. and Galli, S. and Ganga, K. and {G{\'e}nova-Santos}, R. T. and Gerbino, M. and Ghosh, T. and {Gonz{\'a}lez-Nuevo}, J. and G{\'o}rski, K. M. and Gratton, S. and Gruppuso, A. and Gudmundsson, J. E. and Hamann, J. and Handley, W. and Hansen, F. K. and Herranz, D. and Hildebrandt, S. R. and Hivon, E. and Huang, Z. and Jaffe, A. H. and Jones, W. C. and Karakci, A. and Keih{\"a}nen, E. and Keskitalo, R. and Kiiveri, K. and Kim, J. and Kisner, T. S. and Knox, L. and Krachmalnicoff, N. and Kunz, M. and {Kurki-Suonio}, H. and Lagache, G. and Lamarre, J.-M. and Lasenby, A. and Lattanzi, M. and Lawrence, C. R. and Jeune, M. Le and Lemos, P. and Lesgourgues, J. and Levrier, F. and Lewis, A. and Liguori, M. and Lilje, P. B. and Lilley, M. and Lindholm, V. and {L{\'o}pez-Caniego}, M. and Lubin, P. M. and Ma, Y.-Z. and {Mac{\'i}as-P{\'e}rez}, J. F. and Maggio, G. and Maino, D. and Mandolesi, N. and Mangilli, A. and {Marcos-Caballero}, A. and Maris, M. and Martin, P. G. and Martinelli, M. and {Mart{\'i}nez-Gonz{\'a}lez}, E. and Matarrese, S. and Mauri, N. and McEwen, J. D. and Meinhold, P. R. and Melchiorri, A. and Mennella, A. and Migliaccio, M. and Millea, M. and Mitra, S. and {Miville-Desch{\^e}nes}, M.-A. and Molinari, D. and Montier, L. and Morgante, G. and Moss, A. and Natoli, P. and {N{\o}rgaard-Nielsen}, H. U. and Pagano, L. and Paoletti, D. and Partridge, B. and Patanchon, G. and Peiris, H. V. and Perrotta, F. and Pettorino, V. and Piacentini, F. and Polastri, L. and Polenta, G. and Puget, J.-L. and Rachen, J. P. and Reinecke, M. and Remazeilles, M. and Renzi, A. and Rocha, G. and Rosset, C. and Roudier, G. and {Rubi{\~n}o-Mart{\'i}n}, J. A. and {Ruiz-Granados}, B. and Salvati, L. and Sandri, M. and Savelainen, M. and Scott, D. and Shellard, E. P. S. and Sirignano, C. and Sirri, G. and Spencer, L. D. and Sunyaev, R. and {Suur-Uski}, A.-S. and Tauber, J. A. and Tavagnacco, D. and Tenti, M. and Toffolatti, L. and Tomasi, M. and Trombetti, T. and Valenziano, L. and Valiviita, J. and Van Tent, B. and Vibert, L. and Vielva, P. and Villa, F. and Vittorio, N. and Wandelt, B. D. and Wehus, I. K. and White, M. and White, S. D. M. and Zacchei, A. and Zonca, A.},
  year = {2020},
  month = sep,
  journal = {Astronomy \& Astrophysics},
  volume = {641},
  eprint = {1807.06209},
  primaryclass = {astro-ph},
  pages = {A6},
  issn = {0004-6361, 1432-0746},
  doi = {10.1051/0004-6361/201833910},
  urldate = {2022-12-04},
  archiveprefix = {arXiv}
}

@article{riessComprehensiveMeasurementLocal2022,
  title = {A {{Comprehensive Measurement}} of the {{Local Value}} of the {{Hubble Constant}} with 1 Km/s/{{Mpc Uncertainty}} from the {{Hubble Space Telescope}} and the {{SH0ES Team}}},
  author = {Riess, Adam G. and Yuan, Wenlong and Macri, Lucas M. and Scolnic, Dan and Brout, Dillon and Casertano, Stefano and Jones, David O. and Murakami, Yukei and Breuval, Louise and Brink, Thomas G. and Filippenko, Alexei V. and Hoffmann, Samantha and Jha, Saurabh W. and Kenworthy, W. D'arcy and Anand, Gagandeep and Mackenty, John and Stahl, Benjamin E. and Zheng, Weikang},
  year = {2022},
  month = jul,
  journal = {The Astrophysical Journal Letters},
  volume = {934},
  number = {1},
  eprint = {2112.04510},
  primaryclass = {astro-ph},
  pages = {L7},
  issn = {2041-8205, 2041-8213},
  doi = {10.3847/2041-8213/ac5c5b},
  urldate = {2023-04-19},
  archiveprefix = {arXiv}
}

@article{saadehHowIsotropicUniverse2016,
  title = {How Isotropic Is the {{Universe}}?},
  author = {Saadeh, Daniela and Feeney, Stephen M. and Pontzen, Andrew and Peiris, Hiranya V. and McEwen, Jason D.},
  year = {2016},
  month = sep,
  journal = {Physical Review Letters},
  volume = {117},
  number = {13},
  eprint = {1605.07178},
  primaryclass = {astro-ph},
  pages = {131302},
  issn = {0031-9007, 1079-7114},
  doi = {10.1103/PhysRevLett.117.131302},
  urldate = {2025-06-08},
  archiveprefix = {arXiv}
}

@article{saidJointAnalysis6dFGS2020,
  title = {Joint Analysis of {{6dFGS}} and {{SDSS}} Peculiar Velocities for the Growth Rate of Cosmic Structure and Tests of Gravity},
  author = {Said, Khaled and Colless, Matthew and Magoulas, Christina and Lucey, John R and Hudson, Michael J},
  year = {2020},
  month = sep,
  journal = {Monthly Notices of the Royal Astronomical Society},
  volume = {497},
  number = {1},
  pages = {1275--1293},
  issn = {0035-8711, 1365-2966},
  doi = {10.1093/mnras/staa2032},
  urldate = {2025-05-29},
  copyright = {https://academic.oup.com/journals/pages/open\_access/funder\_policies/chorus/standard\_publication\_model}
}

@article{soltisPercentLevelTestIsotropic2019,
  title = {Percent-{{Level Test}} of {{Isotropic Expansion Using Type Ia Supernovae}}},
  author = {Soltis, John and Farahi, Arya and Huterer, Dragan and Liberato, C. Michael},
  year = {2019},
  month = mar,
  journal = {Physical Review Letters},
  volume = {122},
  number = {9},
  pages = {091301},
  issn = {0031-9007, 1079-7114},
  doi = {10.1103/PhysRevLett.122.091301},
  urldate = {2024-08-01}
}
\end{document}